\documentclass{edm_template}
\usepackage{float}

\usepackage{tabularx, colortbl}
\usepackage{listings}
\usepackage{algpseudocode}
\usepackage{booktabs}
\usepackage{rotating}
\usepackage{multirow}
\usepackage{adjustbox}
\usepackage{graphicx}
\usepackage[stable]{footmisc}
\usepackage{hyperref}
\usepackage{makecell}
\usepackage{subcaption}
\usepackage{amsmath}
\usepackage{amssymb}
\usepackage{enumitem}

\usepackage{balance}
\usepackage[subtle]{savetrees}
\begin{document}

\title{What will you do next?\\A sequence analysis on the student transitions between online platforms in blended courses
}


\numberofauthors{4} 
\author{
\alignauthor
Niki Gitinabard\\
      \affaddr{North Carolina State University}\\
      \affaddr{Raleigh, NC, USA}\\
      \email{ngitina@ncsu.edu}
\alignauthor
Tiffany Barnes\\
      \affaddr{North Carolina State University}\\
      \affaddr{Raleigh, NC, USA}\\
      \email{tmbarnes@ncsu.edu}
\alignauthor
Sarah Heckman\\
      \affaddr{North Carolina State University}\\
      \affaddr{Raleigh, NC, USA}\\
      \email{sarah\_heckman@ncsu.edu}
\and
\alignauthor
Collin F. Lynch\\
      \affaddr{North Carolina State University}\\
      \affaddr{Raleigh, NC, USA}\\
      \email{cflynch@ncsu.edu}
}

\maketitle
\begin{abstract}
    Students' interactions with online tools can provide us with insights into their study and work habits. Prior research has shown that these habits, even as simple as the number of actions or the time spent on online platforms can distinguish between the higher performing students and low-performers. These habits are also often used to predict students' performance in classes. One key feature of these actions that is often overlooked is how and when the students transition between different online platforms. In this work, we study sequences of student transitions between online tools in blended courses and identify which habits make the most difference between the higher and lower performing groups. While our results showed that most of the time students focus on a single tool, we were able to find patterns in their transitions to differentiate high and low performing groups. These findings can help instructors to provide procedural guidance to the students, as well as to identify harmful habits and make timely interventions.
    
\end{abstract}

\section{Introduction}

Modern blended classrooms are defined by suites of educational tools such as learning management systems, online forums, intelligent textbooks, video lectures, groupware tools, and even ticketing systems for office hours. The ubiquity of such tools provides researchers with a rich amount of data on students' study behaviors, work habits, and their learning trajectories. This data can help researchers to identify good and bad study habits among students as well as to define measures for estimating students' performance early on in the courses.  Large datasets of this type first became available in Massive Open Online Courses (MOOCs) that have supported informative research on students' online study habits.  While these tools have now become the norm in many classroom settings and while there has been substantial research on how students use the individual tools, we have far less understanding of how students work across tools and how different patterns of use may affect their learning. Our goal in this research is to address this question through the use of sequence mining. By developing a better understanding of student activities in different online systems and their transitions between these tools, we can provide the instructors with insight on how their students usually behave when they are not in class. 

Prior research has shown that there are several features easily extracted from user logs that can distinguish high performing students from the lower performing ones. Researchers have found several informative features such as number of videos watched per week, completing assignments \cite{pursel16}, starting early \cite{rose14, yang13}, or skipping videos and assignments \cite{halawa14} that were associated with students' performance and dropout in MOOCs.  Studies in blended courses showed that features such as course attendance, web page views, number of watched videos, number of pauses in videos, and the number of attempts before getting each question right are correlating with student dropouts \cite{chen17}. 

More recent work in MOOCs, Intelligent Tutoring Systems (ITSs), and blended courses has focused on grouping the student activities into study sessions and analyzing these sessions and the sequence of students' actions in them. Some researchers have analyzed features based upon these sessions in MOOCs and blended courses, such as the duration \cite{amnueypornsakul14, sheshadri18}. However, those studies overlook the patterns of student transitions between different states or different tools. Other researchers have studied the sequences of student actions in each session, but most of those studies are focused on MOOCs or ITSs and not many of them have focused on blended courses and the data collected from the several tools that the students use for these classes. Some of these studies have relied on Hidden Markov Models on the sequences of student actions and compared the diagrams between high and low performing students (e.g. \cite{faucon16, geigle17, jeong08}), while others have clustered these sequences to find groups of similarly behaving students in classes (e.g. \cite{boroujeni18, guerra14, desmarais13, kizilcec13, klingler16, kock11, patel17, shih10}). These studies have often been able to identify relevant clusters among the students such as ``confirmers'' and ``non-confirmers'' \cite{guerra14} or ``behind'', ``on-track'', ``auditing'', and ``out'' \cite{kizilcec13}. Also, other sets of studies have performed differential pattern mining on such sequences to find the patterns that are different between high and low performing students \cite{kinnebrew13, kinnebrew12, herold13, mukala15}. And finally, another part of this research treats the sequences of actions as strings and uses analysis of N-grams to identify the popular trends in student activities and transitions \cite{li17, Brooks2015, sinha14, wen14}. These methods are helpful in revealing many of the students' behavioral patterns and the differences between the different performance groups, but are mostly focused on MOOCs or ITSs.

Despite the extensive research in this area on MOOCs and ITSs, studies on student transactions in blended courses are limited and most of them focus on correct/incorrect attempts on the same platform (e.g. the assignment submission systems) \cite{guerra14}. In this work, we collected activity logs from four online platforms for two offerings of two on-campus classes at North Carolina State University. In these classes, Piazza was used as a discussion forum, Moodle as a Learning Management System (LMS) was the means of sharing the course material and assignments, Github was used in one class as a version control as well as a code submission tool for the projects, and WebAssign was used for assignment submissions and automated grading in the other class. We aligned the logs into a single coherent transaction record, grouped the individual student actions into study sessions, and extracted the sequences of student actions from them. Finally, we labeled the students as the ``Distinction'' group who gained an A- or above and the ``Non-distinction'' group who gained a B+ or below in these courses and used N-gram analysis as well as Apriori studies to find the answers to the following research questions:

\begin{enumerate}[label=\textbf{RQ\arabic*}]
    \item What are the most common transitions between different course tools?
    \item Which transitions are significantly different between the distinction and non-distinction groups?
   
\end{enumerate}

The answers to these questions can help us understand the trends of student activities better, to find key differences between high-performing students and the lower performing ones, and help the instructors to provide guidance to the students as they work or identify harmful patterns early in the semesters.

\section{Literature Review}
\subsection{Students' Online Activity Analysis}

Since detailed online student logs have been available for the MOOCs, there have been extensive studies of student behaviors using these logs to identify their association with the students' performance and attrition. Even simple measures such as number of videos watched are shown to be predictive of students' attrition and performance in MOOCs. Some examples of these features include the number of videos watched per week, whether the student watched all of the lectures, or completed all of the assignments \cite{pursel16}.  They also included joining the course early \cite{rose14, yang13}, skipping videos or assignments, assignment performance \cite{halawa14}, spending more time on each assignment \cite{andres16}, the number of lecture views/downloads, quiz attempts, and forum views/posts/comments \cite{fei15}. Some researchers such as Yang et al. have gone further and constructed more complex features to represent student confusion and shown that increased confusion is associated with dropout in MOOCs \cite{yang16}.  Chen et al. has studied blended courses and has also shown that features such as course attendance, web page views, videos watched, video pauses, and assignment attempts are also correlated with student dropout \cite{chen17}. All of these features, while informative, overlook an important part of the information that online logs provide us: the sequences of actions and transitions among different platforms.

To analyze a group of student actions as a whole, researchers have suggested defining study sessions. Prior work has suggested different methods for defining study sessions such as having a ``fixed duration'' \cite{Brooks2015}, using ``browser navigations'', or having a `cutoff' \cite{amnueypornsakul14}. But as Kovanovic et al. showed, the choice of the method or the cutoff time is not trivial and there is no best method for everyone \cite{Kovanovic:Penetrating:2015}. They suggested exploring the data to find the cutoff or method that matches the dataset best.  Amnueypornsakul et al. defined study sessions and used the actions and the sessions to calculate measures such as the length of the action sequence, the number of occurrences of each activity, and the number of Wiki page views \cite{amnueypornsakul14}. Sheshadri et al. also defined study sessions based on the time difference between student actions and extracted measures such as the average number of actions in each session, inconsistency of the student (i.e. how different the number of the sessions started by a student is from the class average and how infrequent they get online), average length of sessions, and sessions including discussion forum activity \cite{sheshadri18}. While these features can add to the information collected directly from the online tools, they still do not consider transitions from one type of action to the other.

\subsection{Sequence Analysis}
\subsubsection{Markov Models}

Several methods have been used for analyzing the sequences of student actions. The first and most popular is the use of Markov chains and Hidden Markov Models. Jeong et al. for example, trained models based upon system logs of a learning-by-teaching system called Betty's Brain in which the students learn material by teaching an artificial agent, Betty \cite{jeong08}. The possible student actions in this platform are reading the material they are trying to teach Betty; editing the material; using links and concepts in forms of adding, removing, or changing (e.g. link add); querying the agent by asking questions about the provided material; asking Betty the agent to explain the answer she just gave; and giving a quiz to assess how well Betty has learned. The authors extracted sequences of student actions on the platform and used a Hidden Markov Model to analyze their behavior. They found that students who generated better concept maps used balanced learning strategies that include moving between different actions, while the students who generated low scoring concept maps typically focused too much on getting the quiz answers correct. Faucon et al. used semi-Markov chains to model student activities in 61 MOOCs offered by EPFL university on Coursera and EdX platforms \cite{faucon16}. They utilized an Expectation Maximization algorithm for fitting the model and showed a graphical representation of their results on the transitions between different states (e.g. submission, forum participation, video watching, etc) for students of different behavior profiles.  Similarly, Geigle et al. used clickstream data from a UIUC Text Retrieval MOOC on Coursera to generate a transition diagram between the different tools \cite{geigle17}. While Markov models are suitable for modeling student transactions between different states and are easy to visualize, the differences they show are often hard to quantify and compare between groups \cite{jeong08}.

\subsubsection{Sequence Clustering}
Another approach for analyzing sequences of student actions is by clustering them. Desmarais et al. for example, collected the action logs of students in a college math learning environment \cite{desmarais13}. In that work, they defined distinct sessions where the students paused for more than 5 minutes between them unless the action after the pause was a submission to an exercise, which might take longer. They then clustered the sequences using the Levenshtein distance and identified three types of sessions. The first was when the students showed exploratory behavior and engaged in a mixture of browsing through exercises and notes. The second type were the short sessions comprising a variety of behaviors such as browsing and attempting the exercises and quizzes. The third were exercise intensive sessions mostly consisting of exercise logs. Kizilcec et al. used a similar approach on the student engagements in a MOOC \cite{kizilcec13}. For each assessment period, they labeled the students as either ``behind'', ``on track'', ``auditing'', or ``out'' based on their engagement with the course material. Then, they applied K-means clustering on the sequences of the student states in all assessment periods to identify the prototypical engagement patterns and were able to observe four clusters of students as completing, auditing, disengaging, and sampling. 

A similar analysis was performed by Guerra et al. on data collected from QuizJET, which was a voluntary practice platform for students in an introduction to programming blended course \cite{guerra14}. They extracted the sequence of correct and incorrect submissions for each student and each question. Then, by comparing the sequences of different students to the sequences of the same student, they observed that these sequences are personal and can show people's study approaches like a ``study genome''. While these genomes were shown to evolve throughout the semester, the evolved genomes for a single user were still more similar than the genomes across different users. They were able to cluster the students based on their genomes and identify two groups as the confirmers and the non-confirmers. The confirmers kept trying examples of the same topic even after they got one correct, while the non-confirmers moved on to the next topic after they were able to solve one example correctly. Finally, Boroujeni et al. clustered student activities in a MOOC and were able to identify four user profile types: users who watch videos before making submissions (44\% of the users), users who make submissions without watching videos (2\% of the users), users who watch videos and never submit (7\% of the users), and the users who change their habit in the semester (47\% of the users) \cite{boroujeni18}. These categories are similar to the ones suggested by Kizilcec et al. \cite{kizilcec13}. While clustering seems to offer much insight on similar sequences and differences between different groups of students, it is often challenging to interpret these clusters and get to real world groups of students.

To account for the randomness in the generation of Markov Models, some researchers have generated Markov Models based upon each individual sequence and then clustered them to obtain more meaningful results. K\"ock et al. for example, designed an analysis pipeline which included a pre-processor which extracted activity sequences from the raw data, a modelling unit which converted the sequences into Deep Markov Models, and a final clustering unit \cite{kock11}. They applied this pipeline to extract common transitions exhibited by different performance groups in a Physics course at the US Naval Academy. Similarly, Shih et al. applied the same clustering method on the Hidden Markov Models based on student activities in a Geometry Cognitive Tutor \cite{shih10}. Klingler et al. developed an evolutionary clustering pipeline to improve cluster stability over multiple training sessions in the presence of noise \cite{klingler16}. This pipeline extracts action sequences from log data, transforms them into per-session Markov Chains, computes pairwise similarities between students for every session, then performs clustering using evolutionary clustering, and uses the Akaike information criterion with correction (AICc) to select the best model. They suggested that this pipeline can be used as a black box on any ITS. While the combination of clustering and Markov Models might overcome some disadvantages of each individual, the results are still challenging to interpret as noted by Shih et al. \cite{shih10}.

\subsubsection{Sequences as N-grams}
Another approach often taken when analyzing students' sequence data is treating the sequence of actions as a sequence of strings, and then identifying the common N-grams in it. Li et al. and Sinha et al. for example, extracted the sequences of actions for users in MOOCs and used the frequency of N-grams in such sequences as predictive features to predict students' performance and certification \cite{li17, sinha14}. Maldonado et al. also performed a similar analysis on data extracted from an interactive tabletop (Digital Mysteries) and were able to identify frequent sequences of actions that distinguish between different performance groups \cite{maldonado11}. Wen and Ros\`e applied this method to extract the most common types of sessions among students and were able to identify 4 types of sessions as lecture and peer assessment sessions, browse course sessions, assignment and forum sessions, final quiz and survey sessions, and lecture and quiz sessions \cite{wen14}. Brooks et al. defined fixed duration sessions during the semester (i.e. 1 day, 3 days, 1 week, and 1 month) and recorded students' activity in each frame as a binary feature \cite{Brooks2015}. They used frequent N-grams extracted from these sequences as features to make early and cross-class predictions of student dropout. While N-grams are easier to process since there are many available libraries for analyzing them, extracting information from them can still be challenging and require expert help at times.

\subsubsection{Differential Pattern Mining}
A newer approach which is mostly applied to ITS data is Differential Pattern Mining. The algorithms in this approach are able to identify patterns that are more frequent than a specific threshold and are significantly different between the two specified groups such as pass/fail students \cite{agrawal93}. Kinnebrew et al. for example used a differential sequence mining algorithm to extract the sequences that are different between the high performers and low performers using the Betty's Brain platform \cite{kinnebrew12, kinnebrew13}. They found that the high performers more frequently engaged in reading activities in a monitoring context, while the lower performers usually perform short reads mostly not relevant to their recent actions \cite{kinnebrew13}. Herold et al. applied the same analysis to the sequences collected with LivescribeTMdigital pens, used to complete all of their homework and exams \cite{herold13}. These pens are able to log students' handwriting as time-stamped pen strokes providing the sequence in which it was written. Using this method, they were able to identify 98 patterns in total and use them to make predictions on the students' performance in the closest exam after the task with an $R^2$ of 0.3. While this approach is able to make the differences in performance groups bolder, it is still relatively new, the libraries for it are limited, and it is also possible to end up with a large number of rules that will need clustering again.

\section{Dataset}
We collected data from two offerings each of two distinct courses, a Discrete Math course (DM) in the Fall semesters of 2013 and 2015, and a Java programming course (Java) in the Fall of 2015 and 2016. The 2015 offerings of these courses occurred contemporaneously. Both of these courses are core undergraduate courses, required for students majoring and minoring in Computer Science. They both use significant online materials and support and can be considered blended courses. The online materials include online assignments, supplemental material, and student forums. 

In all these classes, Moodle is used as an LMS for providing the course material and the assignment descriptions to the students. Piazza is used as the discussion forum and the main resource for the students in these courses to ask questions and get answers from the teaching staff as well as to have discussions with their peers. The students were able to post completely anonymously for a brief time in DM-2013 but it was blocked in all other courses. Posting anonymously to other students was always allowed. Posting on Piazza was not required in any of these classes, but it was encouraged by the teaching staff as the best choice of asking for help. In the DM classes, the instructors used multiple answer questions on WebAssign for a large portion of the assignments. WebAssign was configured to allow the students to attempt each question several times to get it correctly and provides the students with instant automatic feedback on their answers. In the Java classes, the students use Github as a version control for keeping track of their code and editing in teams, as well as the means for submitting their code for grading. The students' Github repositories were connected to Jenkins servers, which ran several test cases on their code after each pushed commit. Some of the tests were predefined and authored by the instructional staff and some others were the tests designed by the students to test their own code. This enabled students to get instant feedback on their code and possibly revise it after each submission. Our datasets in this study consist of the Piazza discussions, Moodle logs, and final grades for all the classes as well as Github commit logs for Java classes and WebAssign logs for DM-2013.

While some of the tools used in these classes are different, they play similar roles in the classes.  In the DM classes students use WebAssign to submit their assignments and to receive immediate automated feedback.  And in this class they can re-submit as many times as they wish to get the right answer.  Similarly, in the Java classes the students use Github for making submissions on their projects. While these submissions often take more time than answering a simple question on WebAssign, the students are still able to get immediate feedback from Jenkins and to try again. Consequently, while some visible trends in these classes might be different, we expect the trends for WebAssign and Github to be similar, because they play a similar role. Similarly, in both these classes, Moodle and Piazza can be considered as support platforms since the students can use the course material, project descriptions, and the questions on the forum to resolve their confusions. The types of support these platforms are offering are quite different, since asking questions on Piazza is a more direct means of asking for help than referring to the class material. 

\begin{figure}[h]
  \centering
  \includegraphics[width=0.47\textwidth]{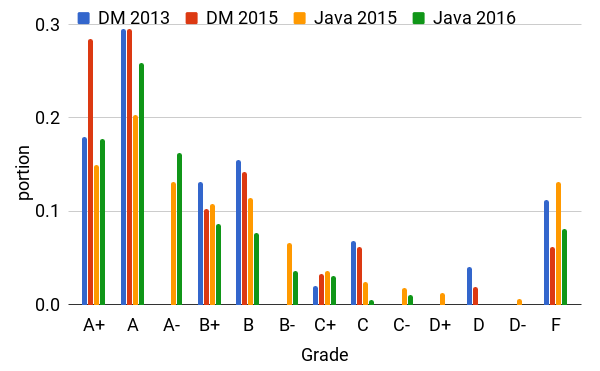}
  \caption{The Distribution of Grades in Different Classes}
  \label{fig:grade_dist}
\end{figure}

More information on the population of these classes is shown in Table~\ref{tab:stats}. The grade distributions for these classes are shown in Figure \ref{fig:grade_dist}. Both these courses are C-wall courses, where the students need a C or better in them to proceed to the next computer science courses in the curriculum. As shown in these figures, most of the students performed well in these classes. Thus, we decided that clustering them into pass/fail groups would be uninformative and result in a skewed dataset.  Since the median grades for all these datasets were close to 90, the cutoff between an A- and a B+ in the courses, we decided to partition the classes into two groups, the \emph{distinction} group earning an A- or above, and the \emph{non-distinction} earning a B+ or below.  This partitioning resulted in an almost even groups of the students. We believe that this segmentation leaves room for adjusting the analysis for other classes with different grade distributions.

\begin{table}
\centering
\caption{Statistics of Each Class}
\label{tab:stats}
\begin{adjustbox}{width=0.48\textwidth}

\begin{tabular}{|l|c|c|c|c|} \hline
\textbf{Class} & \textbf{DM-2013} & \textbf{DM-2015} & \textbf{Java-2015} & \textbf{Java-2016}\\ \hline
Total Students & 251 & 255 & 181 & 206 \\ 
Teaching Assistants & 5 & 5 & 9 & 9 \\
Instructors & 2 & 2 & 4 & 4 \\
Average Grade & 81.2 & 87.6 & 79.7 & 79.9\\
\hline
\end{tabular}
\end{adjustbox}
\end{table}

\subsection{Discrete Math}
This course covered material such as propositional logic, predicate calculus, methods of proof, elementary set theory, the analysis of algorithms, and the asymptotic growth of functions. The total enrollments in these classes consisted of 251 students in DM-2013 and 255 students in DM-2015. Both of these classes were offered in two sections by two instructors with 5 shared teaching assistants. The average final grade in DM-2013 was 81.2 and 87.6 in the 2015 class. Both sections in each year shared the same Moodle page for assignments and class material, a Piazza forum for discussions, and both used WebAssign as well as hand-graded assignments. The only major difference between these two offerings was that in 2015 the instructor consciously delayed responding to posts on Piazza so that the TAs and other students would be more involved. However, most of the posts were still answered in a similar time frame to the ones in 2013 by the lead TA in that class.

\subsection{Java Programming Concepts}

The material of the Java class mainly consisted of software design and testing, encapsulation, polymorphism, inheritance, linear data structures, finite-state machines, and recursion. The total enrollment in these classes was 181 students with an average grade of 79.7 in 2015 and 206 students with an average grade of 79.9 in 2016. 

Both of these classes were offered in two different in-person sections by two separate instructors as well as a distance education section by two other instructors, having a total of four instructors with nine shared teaching assistants. We removed the data for the distance education students from our analysis since they were a much smaller group and differed substantially from the local students who could engage in face-to-face interactions. These classes used Piazza for discussions, Moodle for sharing course materials, Github for working on group projects, and Jenkins for automated code evaluation. 

While the teaching material and the methods were mostly similar across both the offerings, there was a major difference in the lab structures for these classes. Both course offerings included lab sessions. In each session, the students completed a short assignment in a team of three with assistance from the teaching staff. One key difference between the course offerings was in the structure of the lab sessions. In 2015, the labs were conducted in 8 class sessions, thus engaging all of the students and the TAs simultaneously. In 2016 however, students were enrolled in separate lab sessions (approximately 24 students each) with a dedicated TA and participated in 12 lab sessions.  Additionally, in 2015, students continued to work with the same peers for all lab assignments while in 2016, they rotated partners after every four tasks, thus giving them a chance to meet and work with a wider variety of people. 

\section{Methods}
\subsection{Action Sequence Generation}
We began by collecting the logs from Piazza, Moodle, Github, and WebAssign for the courses. Later, we merged them into a single class-level transaction file sorted by time. We then generated study sessions on student activities based on their online transactions as discussed in our prior work \cite{sheshadri18}. 

As Kovanovic et al. suggested, we decided to explore our data to find the best method for generating study sessions \cite{Kovanovic:Penetrating:2015}. Since there was no specific time length for the student sessions in our data, we decided to use a set cutoff time, $m$, for defining the sessions. If two consecutive actions are less than m minutes apart, they belong to the same study session. Otherwise, that session ends and the second activity after m minutes is a start of a new session. We plotted the average time differences between sessions, the total number of sessions, and the average number of activities per session for different cutoff times. These plots showed us two points with major changes that were chosen as the cutoff times for ``study sessions'' and ``browser sessions''. We chose 15 minutes as the cutoff time for browser sessions, which show the times that the students have been online for the entire session. We also chose 40 minutes as the cutoff time for study sessions, which allows the time for the students to go offline for coding or solving problems on paper and get back online.  We used this gap between online actions of the students considering that they often work offline before committing their code to Github or solve a problem on paper before submitting an answer on WebAssign. In this work, we focused on study sessions since they showed more transitions between different platforms. The total number of sessions for each group in each class is shown in Table \ref{tab:ses_count}. In the end, we recorded the sequence of student actions in these sessions for further analysis.

\begin{table}[H]
\centering
\caption{The Total Number of Sessions for Distinction and Non-distinction Groups in Different Classes}
\label{tab:ses_count}
\begin{adjustbox}{width=0.48\textwidth}
\begin{tabular}{|lcc|} \hline
Class Name & Count in Distinction & Count in Non-distinction \\
 \hline
DM-2013 & 7,697 & 6,533\\
DM-2015 & 6,574 & 3,434\\
Java-2015 & 12,219 & 12,786\\
Java-2016 & 19,913 & 9,829\\
\hline
\end{tabular}
\end{adjustbox}
\end{table}

Similar to Kinnebrew et al. and Maldonado et al., we decided to compact the action sequences \cite{kinnebrew13, kinnebrew12, maldonado11}. For that purpose, we replaced consecutive occurrences of the same actions by the ``+'' notion (e.g. MMM was replaced with M+). Our prior work showed that 90\% of the student sessions consisted of access logs to the same platforms \cite{sheshadri18}. Also, the nature of most of these platforms requires consecutive submissions, such as multiple commits to Github for solving issues or multiple submissions on WebAssign until they find the right answer and there is not much of a difference between asking a question on Piazza after 5 submissions or 6. Abstracting these repetitions helps us spot the transitions between these platforms more easily and spot more similar sequences among students. 

\subsection{Sequence Mining}
In order to explain our methods, we first need to define the common terminology in sequence mining. Based on Agrawal et al., the ``support'' of a sequence is defined as the ratio of occurrences of that sequence among all the sequences in the data \cite{agrawal93}. For example, if a sequence S has happened 10 times among a student's study sessions and the student has a total of 100 occurred sequences, the support for S will be 0.1 for that student. Looking at the support metric helps us to look into what percentage of this student's sequences are S, rather than how many occurrences of S this student has. It also simplifies the comparisons between high and low performing students, since generally, the number of all actions for high performing users are higher and this might stop us from spotting the major differences between the students from different performance groups.

Another term often used in sequence mining is `confidence'. Based on Agrawal et al., the confidence of the action B following the action A ($A\rightarrow B$) shows how likely it is for action B to occur after A and is defined as:
$$Confidence(A \rightarrow B)= \frac{Support(A \cap B)}{Support(A)}$$

To identify the most common patterns among the students, we applied the idea of N-gram analysis as in prior work \cite{li17, Brooks2015, sinha14,wen14,maldonado11}. In text mining, an N-gram of length N (e.g. bigram) refers to a specific sequence of N words. Many times, the frequency or the count of N-grams are calculated and used as features. In this work, we treated the sequences of student actions as lists of words and using Scikit-learn library in Python \cite{scikit-learn}, for each student, we calculated the support for all sequences of lengths of 2 - 3 to represent the transitions between every two tools and also keep room to count for repetitions. Then, we collected these numbers for the distinction and non-distinction groups into two separate lists for each sequence. We extracted the average support percentage for each sequence in each group to find the most common patterns among them. Additionally, we performed Kruskal-Wallis (KW) ANOVA test between the two lists for all sequences to find the patterns that occur with a different distribution among these two groups \cite{kruskal1952use}. The Kruskal-Wallis test is a good choice in this context because it does not assume normally-distributed data.

Also, to determine how likely the students are to transfer to a system after using another, we used the Apriori algorithm provided in Apyori library in Python. This algorithm is used to mine frequent item-sets and association rules \cite{agrawal93}. It takes a minimum required support and performs in an incremental order, starting with single items (i.e. 1-sequences) that meet the support requirement ($L_1$) and add other items to the set as long as the support meets the criteria ($L_k$). In this work, we set the minimum support to a low number (0.02) to be able to find and compare even the rare transitions and the 1-sequences were defined as single actions on each platform. Based on Agarwal et al. the pseudo-code for this algorithm is as below:\\

\begin{algorithmic}
\State $L_1$ = frequent 1-sequences \\
\For {$(k = 2; L_{k-1} \neq \emptyset; k++)$}
   \State $C_k =$ New candidates generated from $L_{k-1}$\\
    \For{All possible sequences c}
        \State Increment the count of all candidates in $C_k$ that are contained in c 
    \EndFor
    \State $L_k =$ Candidates in $C_k$ with minimum support 
\EndFor
\State Answer = Maximal Sequences in $\bigcup{k}{L_k}$
\end{algorithmic}

To find the transitions often associated together, we applied the Apriori algorithm on the sequences from distinction students and non-distinction students and calculated confidence for the frequent ones.

While participation on Piazza was not mandatory, it was strongly encouraged by the instructors as the primary venue for help seeking in all of the courses. As a result, we would expect to observe a large number of transitions between the submission tools (i.e. WebAssign and Github) and Piazza. We also expect these transitions to be more frequent after students make consecutive submission attempts since students who struggle with assignments often make several tries before contacting the instructors. We also expect higher-performing students to make more of such transitions because seeking help when they are struggling, rather than postponing it for later or going without, will help them to perform better in the course.

\section{Results}
Since the tools used in these systems are different, we will present our results in each part for each class separately.

\subsection{RQ1. What are the most common transitions between different course tools?}

\subsubsection{DM-2013}
Our prior study on this class had shown that 90\% or more of the student sessions are focused on a single tool and the sessions consisting of all WebAssign actions was the most common across them \cite{sheshadri18}. As Table \ref{tab:seqDM13}, shows, consistent with our prior work, the most common sequence for both performance groups is repeated WebAssign Submissions, covering on average 70\% of action sequences. This is not surprising due to the fact that the students had unlimited submissions on this platform and often sought to ``brute force'' the answers. 

The next most frequent pattern in both groups is multiple Moodle actions, which is again a unsurprising as students are required to log in on each session and must often navigate to their desired resources through a series of actions. Interestingly, transitions between WebAssign and Moodle are also comparatively frequent (the most frequent kind of transition between tools), consisting of approximately 4\% of the total sequences. The more common transitions would be some submissions on WebAssign and moving to Moodle, while this sequence sometimes gets repeated several times as students move between these two tools and we can observe sequences like ``w+m+w'' on average in 0.4\% of the students' transitions or even more complicated ones such as ``m+w+mw+''. Such transitions show students moving between class material like slides and the assignments and may show them referring to slides to revise their answers on WebAssign. We need to note that the sequences longer than 3 actions were not counted towards the calculation of support and confidence and thus, are not shown in the tables. 

One would expect struggling students to move between WebAssign and Piazza to ask questions about the submissions, but as our results show, this transition does not happen frequently. Even among better performing students, it is more common to go to Moodle than Piazza after a couple of submissions, but it is even less likely for the lower-performing students. It seems like the students prefer to find the answers to their confusion among class material rather than asking questions or they prefer to leave help-seeking for another session.

\begin{table}[H]
\centering
\caption{The Support for the Most Frequent Sequences in DM-2013 (W = WebAssign, M = Moodle, P = Piazza) }
\label{tab:seqDM13}
\begin{adjustbox}{width=0.48\textwidth}
\begin{tabular}{|lcc|} \hline
 & Avg in Distinction & Avg in Non-distinction \\
 \hline
W+ & 0.7064 & 0.7227\\
M+ & 0.1408 & 0.1615\\
W+M, M+W, MW, WM & 0.0429 & 0.0380\\
P+  & 0.0303 & 0.0133\\
P+W, W+P, PW, WP & 0.0039 & 0.0006\\
P+M, M+P, PM, MP & 0.0004 & 0.0001\\
\hline
\end{tabular}

\end{adjustbox}
\vspace{-0.35cm}
\end{table}

To better understand the student transitions between WebAssign and Moodle or WebAssign and Piazza, we calculated the confidence score for sequences in which Moodle and Piazza actions occur in the same  session after one or more WebAssign actions. The results of the Apriori algorithm for this class are shown in Table \ref{tab:confDM13}. As we can see, there is almost a 10\% chance of the students going to Moodle after one or more WebAssign submissions, while there is less than a 1\% chance of them going to Piazza.

\begin{table}[h]
\centering
\caption{Confidence for Different Transitions from WebAssign in DM-2013 (W = WebAssign, M = Moodle, P = Piazza)}
\label{tab:confDM13}
\begin{adjustbox}{width=0.32\textwidth}
\begin{tabular}{|lcc|} \hline
 & Distinction & Non-Distinction\\
  \hline
$ W \rightarrow M$ & 0.11 & 0.09\\
$ W \rightarrow P$ & 0.007 & 0.003\\
$ W+ \rightarrow M$ & 0.11 & 0.09\\
$ W+ \rightarrow P$ & 0.007 & 0.003\\
\hline
\end{tabular}
\end{adjustbox}
\end{table}

\subsubsection{DM-2015}
The most frequent sequences for this class are shown in Table \ref{tab:seqDM15}. Unfortunately, in this class, we do not have access to the WebAssign data. Thus, there were far fewer patterns found in this data than the 2013 class. But as with the prior offering, the Piazza actions seem to be not nearly as common as Moodle actions. Additionally, the transitions between Moodle and Piazza were rare.

\begin{table}[h]
\centering
\caption{The Support for the Most Frequent Sequences in DM-2015  (M = Moodle, P = Piazza)}
\label{tab:seqDM15}
\begin{adjustbox}{width=0.48\textwidth}
\begin{tabular}{|lcc|} \hline
 & Avg in Distinction & Avg in Non-distinction \\
 \hline
M+ & 0.913 & 0.966\\
P+ & 0.081 & 0.034\\
PM, MP, M+P, P+M & 0.003 & 0.000\\
\hline
\end{tabular}
\end{adjustbox}
\end{table}

\subsubsection{Java-2015, Java-2016}
The most frequent sequences for these classes are shown in Table \ref{tab:seqJava}. These classes are similar to DM-2015 in that consequent Moodle actions is the most common sequence with an average about 55-65\% of students' sequences in 2015 and about 70-80\% of the student sequences in 2016. While in these classes Github commits are similar to WebAssign activities in DM-2013, the findings show that the students tend to commit their changes far less frequently than they submit questions on WebAssign. Multiple commits on Github are the next most frequent and they occur in about 20-30\% of the student sequences in 2015 and 10-15\% of sequences in 2016. Similar to DM-2013, where the students often moved between the submission system and the course material on Moodle, in these classes 4-6\% of the student sequences are moving between Github and Moodle, where only 0.1-0.5\% of the sequences refer to moving between Github and Piazza. In these classes also, moving back and forth a few times between the platforms is observed and we can see sequences such as ``g+m+g+m'' or ``g+mg+m+''.

\begin{table}[h]
\centering
\caption{The Support for the Most Frequent Sequences in Java Classes (G = Github, M = Moodle, P = Piazza)}
\label{tab:seqJava}
\begin{adjustbox}{width=0.48\textwidth}
\begin{tabular}{|lcc|} \hline
 & Avg in Distinction & Avg in Non-distinction \\
 \hline
& \textbf{Java 2015} &\\
 
M+ & 0.566 & 0.655\\
G+ & 0.294 & 0.204\\
G+M, M+G, GM, MG & 0.041 & 0.043\\
P+ & 0.011 & 0.010\\
P+M, M+P, MP, PM & 0.004 & 0.003\\
P+G, G+P, PG, GP & 0.003 & 0.003\\
\hline
&\textbf{ Java 2016 }& \\

M+ & 0.698 & 0.782\\
G+ & 0.134 & 0.089\\
G+M, M+G, GM, MG & 0.062 & 0.052\\
P+ & 0.012 & 0.005\\
P+G, G+P, PG, GP & 0.005 & 0.001\\
P+M, M+P, MP, PM & 0.003 & 0.002\\
\hline
\end{tabular}
\end{adjustbox}
\end{table}

As with the DM-2013 class, we calculated the confidence score of action sequences that include Moodle and Piazza in the same session after one or more of Github actions. The results of the Apriori algorithm for these two classes are shown in Table \ref{tab:confJava}. As we can see, there is a 28-37\% chance of the students going to Moodle Github submissions, while there is only less than a 3\% chance of them going to Piazza. 

As our results show, the students seem more likely to go to the project descriptions or the course material after some submissions on Github rather than the discussion forum.

\begin{table}[h]
\centering
\caption{Confidence for Different Transitions from Github in Java classes (G = Github, M = Moodle, P = Piazza)}
\label{tab:confJava}
\begin{adjustbox}{width=0.32\textwidth}
\begin{tabular}{|lcc|} \hline
 & Distinction & Non-Distinction\\
  \hline
 & \textbf{Java 201}5 & \\
$G \rightarrow M$ & 0.31 & 0.36 \\
$G \rightarrow P$ & 0.03 & 0.03 \\
$G+ \rightarrow M$ & 0.31 & 0.37 \\
$G+ \rightarrow P$ & 0.03 & 0.03 \\
\hline
&\textbf{ Java 2016}  & \\
$G \rightarrow M$ & 0.28 & 0.31\\
$G \rightarrow P$ & 0.02 & 0.007\\
$G+ \rightarrow M$ & 0.32 & 0.34\\
$G+ \rightarrow P$ & 0.02 & 0.01\\
\hline

\end{tabular}
\vspace{0.3 pt}
\end{adjustbox}
\end{table}


\subsection{RQ2. Which transitions are significantly different between the distinction and non-distinction groups?}

\subsubsection{DM-2013}
The KW p-value results for the support percentages of different sequences in DM-2013 class is shown in Table \ref{tab:KWDM13}. The significant values with $p < 0.05$ are marked as bold, while edge cases with $p < 0.1$ are marked in italics. We only included the significant and edge-case patterns and the transitions between platforms in the table. As these results show, the distinction students are significantly more likely to have a sequence of Piazza actions than the non-distinction group, with an average of 3\% of their activities in the distinction group compared to 1\% in the non-distinction group. The distinction students are also more likely to go to Piazza after a repetition of other activities than the non-distinction group. While the transition between WebAssign and Moodle (W+M, WM, M+W, MW) is high in both groups and not significantly different, the distinction group is more likely to move between Piazza and WebAssign (PW, WP, W+P, P+W) on average 0.4\% compared to 0.01\%.

\begin{table}[h]
\centering
\caption{KW p-values between distinction and non-distinction students for different sequence supports in DM-2013 (W = WebAssign, M = Moodle, P = Piazza)}
\label{tab:KWDM13}
\begin{adjustbox}{width=0.48\textwidth}
\begin{tabular}{|lccc|} \hline

N-gram & Avg in Distinction & Avg in Non\_distinction & KW pvalue \\
\hline
+P & 0.0018 & 0.0001 & \textbf{3.31E-03} \\
P-W transitions & 0.0039 & 0.0006 & \textbf{3.08E-03} \\
P+ & 0.0303 & 0.0133 & \textbf{1.30E-05} \\
M-W transitions & 0.0429 & 0.0380 & 0.644608 \\
\hline
\end{tabular}

\end{adjustbox}
\end{table}

\subsubsection{DM-2015}
The KW p-values for the different sequences between the distinction and non-distinction group are shown in Table \ref{tab:KWDM15}. Similar to the previous offering, the distinction group in this class are also more likely to have consequent Piazza activities, as well as go to Piazza after consequent actions on another platform. They are also more likely to move between Moodle and Piazza, while the non-distinction group is more likely to perform consequent actions on Moodle.

\begin{table}[h]
\centering
\caption{KW p-values between distinction and non-distinction students for different sequence supports in DM-2015 (M = Moodle, P = Piazza)}
\label{tab:KWDM15}
\begin{adjustbox}{width=0.48\textwidth}
\begin{tabular}{|lccc|} \hline

N-gram & Avg in Distinction & Avg in Non\_distinction & KW pvalue \\
\hline
P-M transitions & 0.003 & 0 & \textbf{1.93E-03}\\
+P & 0.001 & 0 & \textit{5.82E-02}\\
M+ & 0.913 & 0.966 & \textbf{5.80E-05}\\
P+ & 0.081 & 0.034 & \textbf{1.18E-04}\\
\hline
\end{tabular}
\end{adjustbox}
\end{table}

\subsubsection{Java-2015}
The KW p-values for the different sequences between the distinction and non-distinction groups are shown in Table \ref{tab:KWJava15}. Similar to the prior classes, the distinction group in this class was also more likely to go to Piazza after consequent actions on other platforms. Also, similar to DM-2015, the transitions between Moodle and Piazza are significantly more likely among the distinction group. Also, the distinction group has significantly more consequent actions on Github than the non-distinction group. However, while on average more sequences have a repetition of Piazza activities among distinction students, this difference is not significant in this class. Similarly, moving between Github and Moodle is more likely on average among the non-distinction group, but this difference is also not significant.

\begin{table}[H]
\centering
\caption{KW p-values between distinction and non-distinction students for different sequence supports in Java-2015 (G = Github, M = Moodle, P = Piazza)}
\label{tab:KWJava15}
\begin{adjustbox}{width=0.48\textwidth}
\begin{tabular}{|lccc|} \hline

N-gram & Avg in Distinction & Avg in Non\_distinction & KW pvalue \\
\hline
M+ & 0.566 & 0.655 & \textbf{0.046}\\
P-M transitions & 0.004 & 0.003 & \textbf{0.022}\\
+P & 0.003 & 0.002 & \textit{0.052}\\
P+ & 0.011 & 0.010 & \textit{0.095}\\
G+ & 0.294 & 0.204 & \textbf{0.005}\\
G-M transitions & 0.041 & 0.043 & 0.788\\
P-G transitions & 0.003 & 0.003 & 0.336\\\hline
\end{tabular}
\end{adjustbox}
\end{table}

\subsubsection{Java-2016}
The KW p-values for the different sequences between the distinction and non-distinction groups are shown in Table \ref{tab:KWJava16}. Similar to the previous classes, in this class also we observe more repetitions of Piazza activities in the distinction group as well as more Piazza activities after a repetition of activities on other platforms. Also, similar to the 2015 Java offering and DM-2015, the non-distinction group is more likely to have consequent actions on Moodle. Despite the other classes, transitions between Github and Moodle as well as Github and Piazza are significantly different in this class and more likely for the distinction group. Comparing these results to the ones in Table \ref{tab:confJava}, the findings seem conflicting since the non-distinction group is more likely to have Moodle activity in the same session after Github activities. However, we need to note that the Apriori algorithm, unlike N-grams, calculates the possibility of Moodle actions occurring after, but not necessarily consequently after, the Github activities. So, it seems like that the non-distinction group are more likely to move to Moodle at some point of the session after Github activities, but less likely to do so consequently after the Github actions.

\begin{table}[H]
\centering
\caption{KW p-values between distinction and non-distinction students for different sequence supports in Java-2016 (G = Github, M = Moodle, P = Piazza)}
\label{tab:KWJava16}
\begin{adjustbox}{width=0.48\textwidth}
\begin{tabular}{|lccc|} \hline

N-gram & Avg in Distinction & Avg in Non\_distinction & KW pvalue \\
\hline
M+ & 0.6976 & 0.7822 & \textbf{1.30E-05}\\
+P & 0.0021 & 0.0010 & \textbf{1.97E-02}\\
P+ & 0.0123 & 0.0048 & \textbf{1.12E-03}\\
G-M transitions & 0.0616 & 0.0521 & \textbf{3.98E-02}\\
P-G transitions  & 0.0046 & 0.0010 & \textbf{3.20E-04}\\
P-M transitions  & 0.0026 & 0.0023 & 0.53\\
\hline
\end{tabular}
\end{adjustbox}
\end{table}

\section{Discussion}
While the classes we analyzed and the offerings within them differ in topic, materials, structure, and instructor approach, our analysis shows that there are common patterns across all of them. 

The first visible pattern is that the students are much more likely to complete consecutive actions on the platform they are already using rather than switching to another platform. In all of the classes, the most common trend is two or more actions on WebAssign followed by Moodle in DM-2013, and Moodle followed by Github in the Java classes, while repetitions of Piazza actions seem to be more rare, even compared to platform switches. This might be due to the fact that most of the activities on Moodle, Github, and WebAssign consist of a sequence of smaller actions. For example, the students are much more likely to solve several problems on WebAssign or attempt a single problem several times, rather than only making a single attempt and leaving the platform. Similarly, on Moodle, the students often need more than one click to reach the material they need to access and on Github, the students are likely to push their code, face a failing test on Jenkins, and make a new commit to solve that issue. However, the actions on Piazza are not as closely monitored. On this platform, only making posts and replies are logged and viewing the posts or replies are not. Thus, the students are much more likely to make a single post or reply without any other visible actions on this platform and that might be a reason why consecutive Piazza actions are not as common as the other tools.

Another common pattern is that in contrast to our expectations, the students in all of the classes were much more likely to go back to the class material and the assignment descriptions on Moodle rather than rely on the discussion forum after one or more tries on their assignments. This was illustrated by the high confidence for transitions from WebAssign and Github (i.e. the submission systems) to Moodle (i.e. the indirect support platform), compared to transitions from these platforms to Piazza (i.e. the direct support platform), even in the higher performing students. As we expected, the visible trends for WebAssign and Github are similar in these classes due to the similarity in their educational role. As mentioned before, since the views are not monitored on Piazza, it is therefore possible that in some cases the students do refer to Piazza posts, only to find their answers in another student's question, without making any posts or replies. Thus, the lower amount of transitions to Piazza might be due to this difference in recording the activities. However, the teaching staff often found that the students did not look for their questions in their peers' posts and kept asking similar questions.

While both the performance groups have a large amount of consecutive Moodle actions, the non-distinction groups have on-average more of such sequences and this difference is often significant in these classes. Also, having repetitive Piazza actions and going back to Piazza after two or more actions on another platform is, on average, more common between the distinction students and this difference is significant in most of the classes, while in other classes an edge case that could be significant if we considered $p < 0.1$. This shows that while the non-distinction group seems to insist on finding the answer among the class material (or possibly reading the existing posts on Piazza), the distinction group seems to ask or answer questions on the discussion forum more often. 

\section{Conclusions}
While multiple researchers have applied sequence analysis to educational data, most of this research has been focused on ITS data or MOOC data and there is not much research on the transitions of students between several resources in blended courses. In this study, we gathered logs from several online platforms that students interacted with in two offerings of two undergraduate courses. We extracted sessions of studies among these activity logs and analyzed the sequences of the student actions in these sessions to find the general patterns in student transitions as well as the patterns that distinguish between the higher performing students and low-performers.

Our results show that consequent actions on the same platform are more likely for the students. Additionally, students are more likely to refer to the class material and the assignment descriptions rather than the discussion forum after a couple of submissions on assignments. However, the higher performers generally had more transitions between platforms and were often more likely to go to the discussion forum than the non-distinction group. We also found that even though some platforms used in classes are different, the results can be generalized across classes as long as the tools play similar educational roles, as WebAssign and Github did in our case. This can help findings to be expanded across a variety of courses using different platforms.

The results of this study can also help instructors identify helpful and harmful patterns among students and offer suggestions for forming more productive habits. The frequencies of these sequences added to the previously defined behavioral features can also help researchers improve the performance of their prediction models on student performances.

One limitation of this study is the differences between the length of the activities and how they are recorded on the different tools. Some types of activities are shorter and thus, more likely to repeat, such as WebAssign submissions where the questions are often multiple answers and quick to submit, while some other activities take a longer time, such as writing a Piazza post or solving an issue with the code and making a new commit. Additionally, while Moodle platform logs every action the users make online, Piazza only records the posts and replies and not the views. These differences in the tools might affect our findings. Further analysis, such as considering the time between actions differently for different tools might help us understand the trends in student activities better. Also, the WebAssign action logs are not available for the DM-2015 class, which limits the findings for this class and makes the comparisons between the two DM offerings less significant. Adding later similar offerings of these courses to the study in the future might help in finding more consistent trends.

In the future, we plan to expand the study to use different sequence analysis tools, such as the differential sequence mining tools. Those tools might be able to highlight other differences among the performance groups that are more difficult to spot using the current tools. Also, replicating our analysis on other courses and more offerings of the same courses can give us a better insight on how general some of these findings are. In the end, we plan on extracting predictive features from the student transitional patterns and add them to the other behavioral features to improve the accuracy of the performance prediction models on students, make the models fit better across classes, or make them fit better for earlier predictions in the semester.

\section{Acknowledgements}
This research was supported by NSF \#1821475 ``Concert: Coordinating Educational Interactions for Student Engagement'' Collin F. Lynch, Tiffany Barnes, and Sarah Heckman (Co-PIs).

\balance
\bibliographystyle{abbrv}
\bibliography{references}
\end{document}